
%

\magnification\magstep 1
\vsize=8.5truein
\hsize=6truein
\voffset=0.5 truecm
\hoffset=1truecm
\baselineskip=20pt

\centerline{{\bf UNCERTAINTY PRINCIPLE AND OFF-DIAGONAL LONG RANGE}}

\centerline{{\bf  ORDER IN THE FRACTIONAL QUANTUM HALL EFFECT}}

\vskip 1.5 truecm

\centerline{L. Pitaevskii$^*$
 and S. Stringari}

\bigskip

\centerline{ {\it Dipartimento di Fisica, Universit\`a di Trento and INFN,
  I-38050 Povo, Italy}}

\vskip 2truecm

{\it {\bf Abstract} A natural generalization of the Heisenberg
uncertainty principle inequality holding for non hermitian operators is
presented and applied to the fractional quantum Hall effect (FQHE).
This inequality was used in a previous paper
to prove the absence of long range order in the ground state of several 1D
systems with continuous group symmetries.
In this letter we
use it to rule out the occurrence of Bose-Einstein condensation in the bosonic
representation of the FQHE wave function proposed by Girvin and MacDonald.
We show that the absence of off-diagonal long range order in this 2D problem
is directly connected with the $q^2$ behavior of the static structure
function $S(q)$ at small momenta.}

\bigskip
\bigskip

{\parindent0pt PACS numbers: 05.30.-d, 73.20.Dx, 73.40.Lq}

\bigskip

\vskip2truecm

$^*${\it permanent address: Kapitza Institute for Physical Problems,
ul. Kosygina 2, 117334 Moscow, Russia}

\vfill\eject

Important results on the role of fluctuations in systems with broken symmetries
have been obtained in the past using a famous  inequality  due to
Bogoliubov$^1$. This inequality provides important constraints on the static
response of the system and yields$^{2,3}$, through the use
of the fluctuation-dissipation theorem, the following result for the
fluctuations of a general physical operator $A$
$$
<\{A^{\dagger},A\}><[B^{\dagger},[H,B]]>\ \ \ge
\ \ k_B T\mid<[A^{\dagger},B]>\mid^2
\eqno(1)
$$
In eq.(1) $A$ and $B$ are two arbitrary (non hermitian in general) operators
and
$H$ and $T$ are, respectively,
 the Hamiltonian and the temperature of the system. Furthermore
$\{A^{\dagger},B\} = A^{\dagger}B + BA^{\dagger}$, \ \
$[A^{\dagger},B] = A^{\dagger}B - BA^{\dagger}$ and  $<>$ means
statistical average (without any loss of generality here and in the following
we assume $<A>=<B>=0$).

Inequality (1) was successfully employed in ref.
[4-6] to prove the absence of long range order at finite temperature in
a relevant class of 1D and 2D systems including Bose superfluids and
superconductors, isotropic ferromagnets and anti-ferromagnets and crystals.
The Bogoliubov inequality is not however particularly useful
in the study of fluctuations in the low temperature regime dominated
by quantum effects. Actually  eq.(1) becomes useless in the zero temperature
limit, while the original Bogoliubov inequality,
without the use of  the fluctuation-dissipation theorem,
provides direct information only on the
static response$^{2,3}$ of the system and not on its fluctuations.

On the other hand fundamental restrictions on quantum
fluctuations are in general provided by the Heisenberg uncertainty principle.
This principle, usually formulated for hermitian operators, can be
naturally generalized to the case of non-hermitian operators starting
from the  inequality
$$
\sqrt{ <A^{\dagger}A> <B^{\dagger} B>} + \sqrt{ <A A^{\dagger}> <B
B^{\dagger}>}
\ \ \ge \ \ \mid<[A^{\dagger},B]>\mid
\eqno(2)
$$
that can be easily proven applying
the Schwartz inequality to the scalar product defined
by $(A,B) \equiv <A^{\dagger}B>$ and using the inequality
$\mid<[A^{\dagger},B]>\mid \ \le \ \mid<A^{\dagger}B>\mid +
\mid<B A^{\dagger}>\mid$.
{}From eq.(2), noting that $\mid a\mid+\mid b\mid \ge
2 \sqrt {\mid a\mid \mid b \mid }$, one immediately obtains the result
$$
<\{A^{\dagger},A\}><\{B^{\dagger},B\}>\ \ \ge \ \ \mid<[A^{\dagger},B]>\mid^2
\eqno(3)
$$
already derived in ref.[7].
When $A$ and $B$ are hermitian both eqs.(2) and (3) coincide  with the usual
uncertainty principle inequality $<A^2><B^2> \ge {1 \over 4}
\mid<[A,B]>\mid^2$.
In ref.[7] we derived result (3) using a different method, based
on auxiliary operators related
to the physical ones through a linear transformation$^8$.
Other generalizations of the uncertainty principle
for non hermitian operators have been considered in the literature (see
for example [9,10]). However, differently from such generalizations, our
inequalities  (2,3) are
characterized by the occurrence of the commutator in the
r.h.s., a typical and important
feature of the traditional Heisenberg uncertainty principle.

It is worth comparing the Bogoliubov inequality (1) with the uncertainty
principle inequality (3). While the former
explicitly accounts for the role of thermal
fluctuations, the latter turns out to be
particularly powerful at low temperatures where quantum fluctuations
become more and more important. Note that result (3) does not involve
 the hamiltonian of the system in an explicit way
and hence it expresses fundamantal
properties of fluctuations, regardless the explicit form of the
interaction and of the energy spectrum of the system.
 Another important feature of result (3) is that
it does not imply statistical equilibrium and can be consequently used by
averaging on arbitrary non equilibrium states.

A useful (also rigorous) inequality, yielding the Bogoliubov (eq.(1))
and the uncertainty principle (eq.(3)) inequalities
in the high and low temperature regimes respectively, is given by
$$
<\{A^{\dagger},A\}><\{B^{\dagger},B\}>\ \ \ge \ \ \mid<[A^{\dagger},B]>\mid^2
cotgh ({\beta \over 2}{<[B^{\dagger},[H,B]]> \over <\{B^{\dagger},B\}>})
\eqno(4)
$$
with $\beta = 1/k_BT$.
This inequality is stronger than (1) and (3) at all temperatures
and belongs to a general class of inequalities than can be derived using,
for example, the formalism of ref.[11].

Several non trivial results
have been recently obtained starting from the uncertainty principle inequality
(3). In particular in ref.[7] we have proven the absence of long range order in
an important class of 1D systems at zero temperature, such as Bose liquids,
isotropic antiferromagnets and crystals. The proof is based on the study
of the infrared
divergent behaviour, induced by a symmetry breaking in the system,
in the fluctuation term  $<\{A^{\dagger},A\}>$ at zero temperature.
These results provide the $T=0$
analogue of the Hohenberg-Mermin-Wagner theorem$^{4-6}$.
Another useful application is the non perturbative study of isospin
impurities in $N=Z$ atomic nuclei$^{12}$, through the explicit determination
of a rigorous lower bound for isospin fluctuations. This is an interesting
problem characterized by a non-spontaneously broken symmetry in a
finite system.

In this work we provide another interesting application of
the unceratinty principle inequality (3) by ruling
out the existence of Bose-like off-diagonal long range order
in the fractional quantum Hall effect (FQHE) at $T=0$.
This result is particularly relevant
because it concerns the absence of long range order in the ground state
of a 2D system.

The bosonic representation of the many body wave function
$$
\Psi_B({\bf r}_1, ..., {\bf r}_N) = exp( {i\over \nu}\sum_{i,j} \alpha_{i,j})
\Psi_F({\bf r}_1, ..., {\bf r}_N)
\eqno(5)
$$
was used in ref.[13] in order to investigate the problem of off-diagonal
long range order
in the FQHE and to point out the existence of deep analogies between
the behaviour of superfluidity and the FQHE.
In eq.(5) $\Psi_F$ is the fermionic wave function
of electrons, $\nu = {1 \over 2k+1}$,
where $k$ is an integer, is the usual filling factor
and $\alpha_{i,j}$ is the angle between the vector connecting
particles $i$ and $j$  and an arbitrary fixed axis.
Using the Laughlin's expression$^{14}$ for the ground state wave function
$\Psi_F$ the authors of ref.[13] concluded that there is not Bose-Einstein
condensation in the bosonic wave function $\Psi_B$,
but only algebraic long range order (see also ref.[15]). The same
result was obtained in ref.[16] starting directly from the
Chern-Simons-Landau-Ginzburg theory (CSLG).

An interesting question is whether the absence  of Bose-Einstein condensation
in $\Psi_B$ follows from the explicit Laughlin's choice for the wave
function $\Psi_F$ (or from corresponding assumptions in the CSLG theory)
or rather has a more general and fundamental reason.
In the  following we will show that the absence of Bose-Einstein condensation
is a direct consequence of the uncertainty
principle inequality (3) applied to a charged system in an external magnetic
field, without  specific assumptions for the ground state wave function
$\Psi_F$. This result differs from
the case of  neutral 2D Bose systems which are instead
expected to exhibit Bose-Einstein condensation at $T=0$.

Let us apply inequality (3) to an arbitrary
Bose system exhibiting Bose-Einstein condensation. We make the choice
$A=a^{\dagger}_{\bf q}$ and $B=\rho_{\bf q}$ where $a^{\dagger}_{\bf q}$
and $\rho_{\bf q} = \sum_{\bf k}a^{\dagger}_{{\bf k}+{\bf q}}a_{\bf k}$
are the usual Fourier components of the
particle creation and density operators
relative to the Bose system. Using the Bose commutation relation
$[a_{\bf q},\rho_{\bf q}] = a_0$, inequality (3) gives:
$(2n_B(q)+1)2NS(q)\, \ge \, \mid <a_0> \mid^2$
where $n_B(q) = <a^{\dagger}_{\bf q}a_{\bf q}>$ and $S(q) = {1\over N}
<\rho_{-{\bf q}}\rho_{\bf q}>$ are the momentum distribution and
the static structure function respectively.
If gauge invariance is broken, the average value $<a_0>$ does not vanish
and its modulus coincides with $\sqrt{Nn_0}$ where $n_0$ is the fraction
of particles in the condensate, a quantity characterizing the long range
order in the system. Inequality (3) then becomes$^{17}$:
$$
n_B(q) \ge {n_0 \over 4S(q)} -{1\over 2} \, \, .
\eqno(6)
$$

Result (6) very explicitly shows the constraints imposed by the uncertainty
principle on the momentum distribution of the system. It is useful to recall
here that use of the Bogoliubov inequality (1), with the same choice for the
operators $A$ and $B$, yields a different
constraint$^4$
$$
n_B(q) \ge {n_0 m k_BT \over q^2} -{1\over 2}
\eqno(7)
$$
useful only at finite temperature. In particular result (7) can be used to
rule out$^4$ the
existence of Bose-Einstein condensation ($n_0 = 0$) in 1D and 2D Bose
systems at $T\ne 0$.

Let us apply result (6)
to the bosonic wave function (5). The following comments are in order
here: i) the static structure functions S(q) relative to the Bose and Fermi
wave functions of eq.(5) are identical, as obviously follows from the
nature of transformation (5); ii) though $n_B(q)$ should not be confused
with the electronic momentum distribution, its normalization is
nevertheless fixed by the total number $N$ of electrons.

The key point to discuss now is  the low $q$ behaviour of the static structure
factor
$$
S(q) = \int S(q,\omega)d\omega
\eqno(8)
$$
where $S(q,\omega)$ is the usual dynamic structure factor. The function
$S(q)$ obeys, at $T=0$, the rigorous inequality
$$
S(q) \le \sqrt{\int \omega S(q,\omega)d\omega \int {1 \over \omega}
S(q,\omega)d\omega} = q \sqrt{{1\over 2M} G(q)}
\eqno(9)
$$
where we have made use of the well known
f-sum rule
$$
\int \omega S(q,\omega) d\omega ={q^2 \over 2m}
\eqno(10)
$$
and introduced the static response function
$$
G(q) = \int {1 \over \omega}S(q,\omega)d\omega \, \, .
\eqno(11)
$$

 In neutral liquids $S(q)$ vanishes linearly with $q$
at zero temperature as a consequence of the finite value
of the compressibility $\chi = lim_{_{q\to 0}} 2mG(q)$ and this behaviour
ensures, in particular, the absence
of long range order ($n_0=0$) in the ground state of 1D Bose systems$^7$.
Charged liquids in an external magnetic field H are characterized by a
suppression of density fluctuations, resulting in the quadratic law
$$
S(q)_{_{q\to 0}} = {q^2 \over 2m\omega_c}
\eqno(12)
$$
for the static structure function where $\omega_c = eH/m$ is the
usual cyclotron frequency (for a discussion of the low $q$ limit
of $S(q)$ in the FQHE see ref.[18]).
Result (12) can be straigthforwardly obtained starting from
the Kohn's theorem$^{19}$ stating that the leading behaviour
of the dynamic
structure function at small $q$ is given by
$$
lim_{_{q\to 0}} S(q,\omega) = {q^2 \over 2m\omega_c} \delta (\omega -\omega_c)
\eqno(13)
$$
and  hence that the cyclotron resonance  exhausts the energy-weighted
sum rule (10). With the assumption that the system has no gapless
excitations$^{20}$ one immediately finds that the same is true also
for the non energy-weighted sum rule (9) and hence one recovers eq.(12).

\noindent
{}From eq.(12) and our inequality (6) we  conclude that $n_B(q)$
diverges as $n_0m\omega_c/2q^2$ and consequently the normalization
condition $\sum_{\bf q} n_B(q) = N$ cannot be fulfilled
in this 2D problem unless $n_0 =0$. The physical interpretation of
this result is very clear (see also ref.[15]):
the magnetic field suppresses the fluctuations of the electronic density
and, according to the uncertainty principle, it increases the bosonic field
fluctuations that destroy the condensate. The logaritmic divergency resulting
from the $1/q^2$ behaviour in $n_B(q)$ emphasizes in an explicit way
the analogies between this problem and the problem of 2D neutral Bose
superfluids at finite temperature . While in the latter case
the absence of long range order
was proven by Hohenberg$^4$ at $T\ne 0$
employing the Bogoliubov inequality (1), yielding result (7) for
the momentum distribution, in this work we have shown
the corresponding  result for the fractional quantum Hall effect at $T=0$.
Of course our results do not exclude algebraic off-diagonal
long range order whose occurrence was explicitly found by Girvin and
MacDonald$^{13}$ and by Zhang$^{16}$,
nor the occurrence of other types of long range
order as recently discussed by Read$^{21}$.

Starting from eq.(13) one can also calculate the static response function
(9) in the low $q$ limit. In the absence of gapless excitations one finds
$$
G(q)_{_{q\to 0}} = {q^2 \over 2M\omega_c^2} \, \, .
\eqno(14)
$$

The vanishing of $G(q)$ for $q\to 0$ expresses the incompressibility
of the system, a peculiar property of the FQHE. The $q^2$ behavior of $G(q)$
is expected to be preserved by the addition of a small amount of impurities
in the system, whose effect is to broaden the cyclotron
resonance (13). Also in this case, using the rigorous inequalities
(6) and (9) one can consequently rule out the presence of long range
order in the system.

\bigskip

Useful discussions with A.Aronov, E.Seiler and S.Zhang are acknowledged.
L.P. wishes to thank the hospitality of the Dipartimento di Fisica
dell'Universit\`a di Trento.

\vfill\eject

{\parindent0pt REFERENCES}

\item{1.} N.N. Bogoliubov, Phys. Abh. SU {\bf 6}, 1 (1962);

\item{2.} H. Wagner, Z. Physik {\bf 195},  273 (1966);

\item{3.} G. Baym, in {\it Mathematical Methods of Solid State
and Superfluid Theory}, eds. R.C. Clark and G.H. Derrick (Oliver
and Boyd, Edinburgh, 1969) p.121;

\item{4.} P.C. Hohenberg, Phys. Rev. {\bf 158}, 383 (1967);

\item{5.} N.D. Mermin and H. Wagner Phys.Rev.Lett. {\bf 17}, 1133 (1966);

\item{6.} N.D. Mermin, Phys. Rev. {\bf 176}, 250 (1968);

\item{7.} L. Pitaevskii and S. Stringari, J. Low Temp. Phys.
{\bf 85}, 377 (1991);

\item{8.} The present derivation of eq.(3) was suggested to us by E. Seiler;

\item{9.} J.M. Luttinger, Suppl. Progr. Theor. Phys. {\bf 37} \&
{\bf 38}, 35 (1966);

\item{10.} V.V. Dodonov, E.V. Kurmyshev and V.I. Man'ko, Phys.
Lett. A {\bf 79}, 150

(1980);

\item{11.} B.S. Shastry, J. Phys. A: Math. Gen. {\bf 25}, L249 (1992);

\item{12.} L. Pitaevskii and S. Stringari, to be published;

\item{13.} S. Girvin and A. MacDonald, Phys. Rev. Lett. {\bf 58}, 1252 (1987);

\item{14.} R.B. Laughlin, Phys.Rev.Lett. {\bf 50}, 1395 (1983);

\item{15.} S.M. Girvin in R. Prange and S.M. Girvin, {\it The
Quantum Hall Effect} (Springer Verlag, 1990) Appendix;

\item{16.} S.C. Zhang, Int. J. Mod. Phys. {\bf 6}, 25 (1992);

\item{17.} Note that the use of  inequality (2) would imply the
stronger constraint
$$
n(q) \ge {n_0 \over 4S(q)} -{1\over 2} + { S(q) \over 4n_0}
$$
for the momentum distribution, valid for $n_0  \ge S(q)$. One can easily
show that this stronger inequality becomes an equality for all values of
$q$ in the Bogoliubov approximation to the dilute Bose gas;

\item{18.} S.M. Girvin, A.H. MacDonald and P.M. Platzman, Phys.Rev.Lett.
{\bf 54}, 581 (1985); Phys. Rev. B {\bf 33}, 2481 (1986);

\item{19.} W. Kohn, Phys.Rev. {\bf 123}, 1242 (1961);

\item{20.} This assumption seems to be necessary in order to prove result (12)
starting from the Kohn's theorem; for alternative derivations of result (12)
see ref.[18];

\item{21.} N. Read, Phys. Rev. Lett. {\bf 62}, 86 (1989).

\bye